\documentclass[12pt]{article}

\begin{document}

\sloppy
\begin{flushright}{SIT-HEP/TM-11}
\end{flushright}
\vskip 1.5 truecm
\centerline{\large{\bf Non-tachyonic brane inflation}}
\vskip .75 truecm
\centerline{\bf Tomohiro Matsuda
\footnote{matsuda@sit.ac.jp}}
\vskip .4 truecm
\centerline {\it Laboratory of Physics, Saitama Institute of
 Technology,}
\centerline {\it Fusaiji, Okabe-machi, Saitama 369-0293, 
Japan}
\vskip 1. truecm
\makeatletter
\@addtoreset{equation}{section}
\def\theequation{\thesection.\arabic{equation}}
\makeatother
\vskip 1. truecm

\begin{abstract}
\hspace*{\parindent}
We consider non-tachyonic hybrid inflation in the context of the 
braneworld cosmology. 
When one considers models for brane inflation, hybrid inflation is a
natural consequence of the tachyon condensation if it appears at the
end of inflation.
In this case, however, reheating is a difficult problem due to the
peculiar properties of the tachyon.
In this paper we show some models for brane inflation where a new type of
hybrid inflation is realized due to the localized matter fields. 
The obvious advantage of our scenario is successful reheating,
which is due to the potential that is localized on the brane.
The serious problem of the loop correction is also avoided.
\end{abstract}

\newpage
\section{Introduction}
\hspace*{\parindent}
In spite of the great success in the quantum field theory, there is still no
consistent scenario in which quantum gravity is included.
The most promising scenario in this direction would be the string
theory.
In the context of the string theory, the consistency is ensured by the
requirement of additional dimensions.
As for the first time, sizes of extra dimensions had been assumed to be
as small as $M_p^{-1}$.
Then later observations showed that there is no reason
to believe such tiny compactification radius\cite{Extra_1}.
In these models with large extra dimensions, 
the observed Planck mass is obtained by the relation $M_p^2=M^{n+2}_{*}V_n$,
where $M_{*}$ and $V_n$ denote the fundamental scale of gravity
and the volume of the $n$-dimensional compact space.
Assuming more than two extra dimensions, $M_{*}$ may be
close to the TeV scale without conflicting any observable bound.
The most natural embedding of this picture in
the string theory context will be realized by the brane construction.
Of course, the brane construction will be a viable candidate for the
Universe even if the fundamental scale is not so low as the TeV scale.
In the brane world scenario,
 there is no obvious reason to 
believe that the fundamental scale is as high as the Planck scale.

Although this new idea inspired many physicists to lead
them to a new paradigm of phenomenology, a drastic modification is
required for the conventional cosmological scenarios.
Models of inflation and baryogenesis\cite{low-baryon} are especially
sensitive to such a low fundamental scale, i.e., $M_* << M_{GUT}$ where
$M_{GUT}$ denotes the standard (old) GUT scale. 
To avoid extreme fine-tuning, one should reconstruct
 conventional scenarios of the standard cosmology.
This requires inclusion of novel ideas that are quite different from the
conventional one.
For example, if one puts the inflaton field on the brane, their masses
are required to be unnaturally small\cite{fine-tune}.
On the other hand, in generic cases, the mass of the inflaton is bounded
from below to achieve successful reheating. 
Thus it seems quite difficult to construct a model for
inflation driven by a field on the brane.
A way to avoid this difficulty is put forward by Arkani-Hamed et
al.\cite{Arkani-inflation}, where inflation is assumed to occur 
before the stabilization of the internal dimensions.
In this case, however, late oscillation of the radion field is a serious 
problem, which may or may not be solved by the second weak
inflation.
One may find other ways to realize inflation in models
with large extra dimensions.
Due to some dynamical mechanisms, the extra dimensions may be stabilized
before the Universe exited from inflation.
If the stabilization of the internal dimensions occurred before the
end of inflation, it is rather difficult to construct a model for
inflation by the field on the brane, since their energy 
densities are suppressed.
In this case one may use the bulk field rather than a field on the
brane\cite{Mohapatra-1, Mazumdar-Bulk-Inflation}.
In this direction, it is found in ref.\cite{matsuda_bulk_inflation} 
that the phase transition of the hybrid inflation 
becomes fast because of the huge 
number of the destabilized Kaluza-Klein modes at the end of inflation.
Although the problem of the slow phase
transition\cite{Mazumdar-Bulk-Inflation} seems to be solved,
another problem appears because of the excited Kaluza-Klein modes.
Overproduction of the excited Kaluza-Klein modes is a serious problem,
because they efficiently emit Kaluza-Klein gravitons when they
decay into lower excited modes\cite{Mohapatra-1}.
Thus one should conclude that the models for bulk inflation with low $M_*$
are not a candidate for the last (including weak) inflation.
There is an another serious obstacle in constructing models for 
hybrid inflation by the bulk field.
Although the scale of inflation is enhanced by the factor 
of $O(M_p^2/M_*^2)$ for the bulk field, 
it is still difficult to produce the required CMB anisotropy,
 because of the serious constraint from the loop
correction\cite{lyth}.
In this respect, realizing successful hybrid inflation with fundamental
scale $M_*$ as low as the TeV scale seems very difficult both for
conventional and weak inflation.
\footnote{To avoid the most serious constraint from the CMB anisotropy,
one may consider curvaton scenario for an alternative\cite{curvaton}.}

In the brane world scenario, 
one may find another interesting possibility,  ``brane
inflation''\cite{brane-inflation}, in which the interbrane distance
is used for the inflaton.
In this case, hybrid version of brane inflation is naturally
obtained by the tachyon instability. 
In generic models for hybrid brane inflation, the system develops
tachyon modes when the brane distance becomes small, then leads
to a natural end of inflation via extra field as in the
 conventional hybrid models.
This type of scenario has been discussed 
on various occasions\cite{tachyonic_brane_inflation, angle_inflation}.
On the other hand, however, there are problems\footnote{The properties
of the tachyon may be {\bf attractive} for the people who want to
explore a new paradigm of the tachyon cosmology\cite{tachyon-quinn}.} 
related to the peculiar properties of the tachyon field.
For example, reheating after inflation is not so easy as in the
conventional hybrid 
 inflation\cite{tachyon_reheating}.
Moreover, the analysis on such a collapsing process requires more
investigations on the quantum deformation of the branes with
manifestation of the couplings to matter fields.
According to the above viewpoints, it should be interesting if one can
construct models for hybrid brane inflation that does not depend on the
physics related to the tachyon.

In this paper we will consider new types of hybrid brane inflation.
The instability at the end of inflation is induced by the localized
 fields on the brane.
During inflation, their cross terms are suppressed by the
brane distance.
Then at the end of inflation, the cross terms become as large as O(1).
We focus our attention to the possible models
where the brane distance drives hybrid inflation after the radion
stabilization.
In our model, unlike other models for hybrid inflation in the brane
Universe, the tachyon is not required.
Inflation ends because the localized fields on each brane
begin to interact.
Then the interaction destabilizes the potential on the brane.
Unlike the tachyon, the field on the brane can reheat the Universe.
The most attractive point in our model is that reheating is realized
by the localized matter fields on the branes, which may have (1)
couplings to the standard model matter fields, while
the serious loop correction to the inflaton mass is well suppressed
during inflation. 

In section 2, we will first make a brief review of hybrid inflation in
the braneworld cosmology.
There are two important possibilities along this direction, one is
hybrid inflation due to the conventional
field\cite{Mazumdar-Bulk-Inflation, 
matsuda_bulk_inflation, hybrid_ippan_on_brane}, and the other is brane
inflation\cite{tachyonic_brane_inflation, angle_inflation}. 
In this paper we consider the latter possibility, inflation due to
the branes at a distance.
Then we show explicit examples of our idea, non-tachyonic hybrid
inflation due to branes separated at a distance.
The potential on one brane is destabilized by the
interaction between fields on the other brane, which terminates the
brane inflation.
The destabilized field can immediately decay into fields in the standard
model. 
The important point is that the field destabilized at the
end of inflation is not a tachyon but a conventional field
on the brane, which may have couplings to the fields in the standard 
model.
In spite of the large interaction after inflation,
the serious loop correction, which is discussed
in ref.\cite{lyth} to put serious constraint on hybrid inflation, is 
exponentially suppressed and negligible during inflation.

\section{Hybrid brane inflation}
\hspace*{\parindent}
If the brane world represents the present Universe, supersymmetry is
required to be broken by soft supersymmetry breaking terms, unless
the fundamental scale is as low as the electroweak scale.
Flat directions, which become flat in the supersymmetric limit, may
have been raised by small supersymmetry breaking terms.  
The brane distance, which will be parametrized by a massless moduli
field in the supersymmetric limit, will also be raised.

Intersecting brane world with angles is discussed by many
authors as a candidate for supersymmetry breaking configurations
that may realize the hybrid brane inflation\cite{angle_inflation}.
Within the context of interesting brane models, it was shown
that obtaining successful inflationary period is possible from slightly 
non-supersymmetric configurations consisting of D4-branes interescting 
at one angle on a six-dimensional compact space\cite{angle_inflation}.
In these models for inflation, the inflaton field is the moduli
field that represents 
the distance between branes, whose potential is raised by
the supersymmetry breaking.
Inflation is terminated by the tachyon condensation that destabilizes the
brane configuration when two branes come close and collapse into lower
dimensional branes.

Although these models are interesting and theoretically attractive, 
reheating and the 
collapsing process is not well investigated, which may have serious 
difficulty.
To clarify our purpose of this paper, we should first discuss why 
reheating is difficult in thess scenarios.
The simplest version of brane inflation begins with a parallel brane and
an anti-brane at some separation.
Although parallel branes preserve supersymmetry and there is 
no force between
them, the brane-antibrane system can break supersymmetry so that a
nonvanishing potential energy appears to induce an attractive force
between them.
This potential drives inflation.
The form of the potential changes once the branes have reached a
critical separation, where branes become unstable to annihilate or
collapse into lower dimensional branes.
The instability of the brane-antibrane system is described by the
condensation of a tachyonic mode\cite{tachon_sen}.
Because of its peculiarity, the tachyon starts from
the unstable maximum at $T=0$ and rolls down to $T\rightarrow \infty$.
Here the crucial difference from a conventional hybrid
inflation is that the tachyon cannot oscillate to reheat the
Universe in the usual way.
Although there are papers in which reheating by varying the tachyon is
discussed\cite{tachyon_reheating}, it is still interesting to
find a model for hybrid brane inflation that is {\bf not} due to the
tachyon condensation.

From the above viewpoints, in order to obtain successful reheating,
we investigate models for hybrid brane inflation without the tachyon.  
In our model, the origin of the supersymmetry breaking is not the angle
between branes.
Here we assume that supersymmetry is broken on the brane. 
We expect that at the end of inflation, when two branes come
close, the trans-bulk interaction between localized fields on different
branes  
becomes O(1) and destabilizes the potential. 
We show two examples in this direction.
In one case the potential is destabilized by a tree level
interaction, and in the other case the destabilization occurs dynamically.

\subsection{Models with tree-level interactions}
\hspace*{\parindent}
Here we consider a brane configuration that does not invoke the tachyon at
the end of brane inflation.
If one considers domain walls instead of branes, walls should consist of 
independent field configurations that satisfy independent BPS conditions
at least tree level.
Supersymmetry is expected to be broken by the fields on the branes.

Let us consider a potential on one brane,
\begin{equation}
V(\phi_a)_{a} = m^2 \phi_a^2 +\lambda_a \phi_a^4
\end{equation}
where the field $\phi_a$ is localized on the brane.
We also consider a potential on the other brane,
\begin{equation}
V(\phi_b)_{b} = \lambda_b \left[ \phi_b^2 - \Lambda^2 \right]^2
\end{equation}
where $\phi_b$ denotes the field that is localized on the brane.
The interaction between $\phi_a$ and $\phi_b$ is exponentiall suppressed
by the brane distance, 
\begin{equation}
\label{potential_int}
V(\phi_a, \phi_b)_{int} = -\lambda_I e^{-(M_* r)^2} \phi_a^2 \phi_b^2
\end{equation}
where $r$ denotes the distance between branes.

In our model, the inflaton field is the moduli field that parametrizes the
distance between branes, which we denote by $\sigma \equiv M_* ^2 r$.
The potential for the moduli field $\sigma$ is flat if the BPS condition
is satisfied, but will be raised when supersymmetry is broken by the
fields on the branes.
The four-dimensional effective potential is 
\begin{equation}
\label{4dpot}
V(\phi_a, \phi_b, \sigma) = m^2 \phi_a^2 +\lambda_a \phi_a^4 +
\lambda_b \left[ \phi_b^2 - \Lambda^2 \right]^2 
-\lambda_I e^{-(\sigma/M_*)^2} \phi_a^2 \phi_b^2
+m_{\sigma}^2 \sigma^2.
\end{equation}
Inflation starts when $\sigma$ is large.
Then the minimum of the potential on each brane is located
at $\phi_a =0$ and $\phi_b=\Lambda$.
Here we assume that the cosmological constant is tuned so that it
vanishes after inflation.

Then what happens at the end of inflation, when $\sigma$ becomes 
smaller than $M_*$ ?  
When two branes come close, the trans-bulk interaction between fields
(\ref{potential_int}) becomes strong, then it destabilizes the potential
$V(\phi_a)_a$.
The destabilization occurs when $m^2 < \lambda_{I} \Lambda^2$.
In this model the localized field $\phi_a$ can couple to the field in
the standard model through O(1) interactions, which makes it easier to 
reheat the Universe after brane inflation.

Our model also presents a new kind of steep binding energy between 
branes or quasi-BPS domain walls.
The localized potential is destabilized only when branes come close,
which appears as a short-range attraction between branes.
The attractive force is induced by the field that breaks
supersymmetry on the brane.

\subsection{Models with dynamical destabilization}
\hspace*{\parindent}
We consider a familiar mechanism for the radiative symmetry breaking
in minimal supersymmetric standard model(MSSM) as the most naive
realization of our idea.
In MSSM,  after supersymmetry breaking, the Higgs potential is
destabilized due to the loop correction from the top quark, because of its 
large Yukawa coupling.
On the other hand, in models of the brane world, one can find many models
for the fermion mass hierarchy 
that utilizes the localization of the matter fields along extra
dimensions\cite{localize_fermion_hierarchy}. 
Here we consider a model in which the top and the Higgs are localized on
different branes, and the supersymmetry breaking is induced by the tree
level soft mass $\sim O(M_*) \sim TeV$. 
In the true vacuum, these two branes must coincide to certificate the
large Yukawa coupling for the top quark.
However, in the early Universe, these branes may be placed at a
distance along extra dimensions, which means that 
 the Higgs potential is not destabilized at this time.
Weak brane inflation can take place within the above settings, if
the effective mass for the brane-distance moduli is well 
suppressed\cite{lyth}.
Reheating is successful in this
model, because the field that decays after inflation is the conventional
Higgs field on the brane. 
The potential near the origin becomes steep for the inflaton field,
because of its exponential dependence on the distance between branes.

In general, inflation model with higher energy scale is more realistic.
To construct such models, we should include additional components so that
their typical scale become higher than the above simplest example.
In ordinary models for inflation, a light inflaton is a problem
since it cannot
reheat the Universe up to the MeV scale.
However, in the above example, the field that decays and reheats the
Universe at the end of inflation may have O(TeV) mass and O(1)
interaction to the fields on the brane.
The mass of the inflaton after inflation becomes much larger than that was
during inflation, because the potential becomes steep near the origin.

\subsection{Cosmological constraints}
\hspace*{\parindent}
When one considers inflation, one of the most obvious expectations will
be that it explains the origin of the cosmic microwave background (CMB)
anisotropy of the present Universe.
On the other hand, sometimes, the requirement from the CMB anisotropy
imposes serious constraint on the models for inflation.
Although the constraint from the COBE measurement disappears when one
considers alternative mechanisms, such as cosmic
strings\cite{vilenkin-book} or curvaton hypothesis\cite{curvaton},
it is still very important to ask whether the inflation itself can
produce the required CMB anisotropy.

For the models that we have discussed above, it is easy to see that no
fine-tuning is required if it is liberated from the COBE constraint.
As we have mentioned above, reheating is successful in our model. 
Thus our models for hybrid inflation are safely used at least for 
weak inflation.

According to the above arguments, here we consider the question whether 
our model can produce the required CMB anisotropy during inflation without
peculiar fine-tunings.

For the standard models for hybrid inflation, the requirement from the
COBE mesurement puts severe bounds on their scales and couplings,
because of the large loop correction.
For example, here we consider the original hybrid inflation
model\cite{original} with the potential 
\begin{eqnarray}
V(\phi, \sigma)&=& V_0 + \frac{1}{2}m_{\sigma}^2 \sigma^2 + 
\frac{1}{2} g \phi^2 \sigma^2 \nonumber\\
&& \frac{1}{4}\lambda \phi^4 - \frac{1}{2} m_{\phi}^2 \phi^2.
\end{eqnarray}
Let us consider the loop correction, which comes from 
the $\phi$ field.
If there is supersymmetry, the result is simplified and only the
logarithmic part of the following form is relevant,
\begin{equation}
\Delta V_{1-loop}(\sigma) \frac{1}{64 \pi^2}
\left( m^4(\sigma) ln \frac{m^2(\sigma)}{\Lambda^2} \right)
\end{equation}
where the effective mass of the $\phi$ field is given by
\begin{equation}
m^2(\sigma) = (g^2 \sigma^2 -m_{\phi}^2)
\end{equation}
and $\Lambda$ is the renormalization scale.
The flatness conditions require\cite{lyth}
\begin{equation}
g \ll \frac{<\phi>}{M_p}
\end{equation}
and the COBE normalization requirement gives an additional constraint
\begin{equation}
<\phi>^4 \sigma_{COBE} \, \ge\,  (10^9 GeV)^5
\frac{V_0^{\frac{1}{2}}}{(1 MeV)^2} 
\end{equation}
where $\sigma_{COBE}$ denotes the expectation value of $\sigma$ when the
COBE scales leave the horizon.
Considering these conditions, one must conclude that hybrid
inflation is not viable at the TeV scale.
In this model, the lower bound for the energy scale is about $10^9$
GeV\cite{curvaton}.
\footnote{In ref.\cite{inverted}, it is discussed that an alternative
model, which is called inverted hybrid inflation, can evade the above
serious constraints and works even at the TeV scale.}

Here we consider the potential (\ref{4dpot}) and examine the loop
correction from $\phi_a$ and $\phi_b$.
In our model, the relevant coupling is given by 
eq.(\ref{potential_int}).
During inflation, when $r$ is much larger than $M_*^{-1}$, the
loop corrections from the field $\phi_b$ does not depend on $\sigma$
because the $\sigma$-dependent part of $m(\sigma)_{\phi_b}$ is actually zero
during inflation.
On the other hand, although the loop corrections from the field $\phi_a$
does not vanish, the $\sigma$-dependent part of $m(\sigma)_{\phi_a}$
is suppressed by the exponential factor in eq.(\ref{potential_int}),
which makes the loop correction irrelevant to the inflaton potential.

Although the serious constraint from the loop correction is removed in
our model, another problem still remains.
Assuming that the inflaton fluctuation is the origin of the structure of
the Universe, one will find the constraint
\begin{equation}
M_p^{-3} \frac{V_0^{3/2}}{V'} = 5.3 \times 10^{-4}.
\end{equation}
This implies that 
\begin{eqnarray}
\sigma_{COBE} &\sim& M_p^{-3}V_0 ^{3/2} (5.3 \times 10^{-4})^{-1}
m_{\sigma}^{-2}\nonumber\\
&\sim& 10^{-4} GeV \left(\frac{V_0}{(10^{5}GeV)^4}\right)^{3/2}
\left(\frac{V_0/M_p^2}{m_\sigma^2}\right)
\end{eqnarray}
where $\sigma_{COBE}$ is is the expectation value of the inflaton 
when scales explored by COBE leave the horizon.
Here we have used the simplest potential for the inflaton field,
$V(\sigma)=V_0 + m_\sigma^2 \sigma^2$.

Obviously, the bare mass for the $\sigma$ field is required to be
smaller than $\sqrt{V_0/M_p^2}$.
If supersymmetry is broken on the brane and the transition to the bulk
fields occurs at the tree level, one can estimate an {\bf upper} limit
for the soft mass by dimensional analysis\cite{Extra_1},
\begin{equation}
m^2_{modulus}\sim G_{4+n_{E}} \frac{|F_{brane}|^2}{R_E^{n_{E}}},
\end{equation}
where $G_{4+n_{E}}$ is the gravitational constant in the $4+n_{E}$
dimensions and $F_{brane}$ denotes the supersymmetry breaking on the
brane.
Without additional symmetries or mechanisms, the soft masses for the
modulus are expected to be a few orders smaller than the above upper limit.
The lower limit will be given by the requirement from the conventional
soft supersymmetry breaking terms in the supersymmetric extension of the 
standard model. 
Then the required supersymmetry breaking on the brane is about 
$F_{brane} \geq$TeV. 
If the mass for the inflaton is required to be much smaller than the
above lower limit,
the model requires peculiar fine-tunings or specific mechanisms
to forbid the soft mass, even if one uses D-term inflation\cite{D-term,
D-term_matsuda}. 
This problem arises when the scale of D-term inflation is much smaller
than the required F-terms on the brane\cite{halyo}.
In our model, the energy density during inflation can be derived from
the D-term, but the scale is assumed to be larger than the TeV scale.
Thus in our case, we may assume $|F_{brane}|^2 < V_0$ during inflation.

To be more explicit, here we consider a model with $V_0 \sim M_*^4
\sim (10^5 
GeV)^4$ and 
$m_\sigma \sim 0.1 \times \sqrt{V_0}/M_p$.  
Then the required value is $\sigma_{COBE} \sim 10^{-2} GeV$,
which is smaller than the expected value of the field when the
trigger field terminates inflation at $\sigma \sim M_*^{-1}$.
As inflation ends at $\sigma \sim 10^5 GeV$, the produced CMB anisotropy
becomes much smaller than the requirement from the COBE mesurement.
Thus we conclude that the fluctuation of the inflation in our model
cannot produce the required structure of the Universe, if the scale of
inflation is as low as the TeV scale.

In our model, the energy density on the brane has a limit $V_0 < M_*^4$,
where $M_*$ is the fundamental scale.
On the other hand, $\sigma_{COBE}$ is bounded from below, because
inflation is terminated 
at $\sigma \sim M_*$.
Considering the above limits, we can find a bound for the fundamental
scale in order to satisfy the requirement from the COBE mesurement.
After simple calculations, we find 
$M_* >10^{6}$ GeV, which is weaker than the one for the 
original model for hybrid inflation.

Our conclusion in this section is the following.
Although the serious constraint from the loop
correction is removed in our model, it is still difficult to produce the
required CMB anisotropy by inflation with low energy scale.
The requirement is $M_* >10^{6}$ GeV, which is an improvement from 
the original model for hybrid inflation.

\section{Conclusions and Discussions}
\hspace*{\parindent}
In this paper, we have proposed new idea for brane inflation, which does
not utilize the tachyon. 
Our model may as well be considered as a novel realization of hybrid 
inflation, which is not discussed yet.

In any models for brane Universe, it is natural to think that some
fields are localized on branes at a distance.
It is also natural to expect that these fields may have
$O(1)$ couplings when branes are on top of each other.
If the interaction destabilizes a potential, our idea for hybrid brane
inflation works.

With these things in mind, we have constructed two explicit examples.
There are two advantages compared to the previous models for 
tachyonic brane inflation or standard hybrid inflation.
The most attractive point is that reheating is natural in our model.
It is also attractive that the serious constraint from the loop
correction is evaded. 
As we have stated above, our settings are quite natural in models for 
the brane world.

\section{Acknowledgment}
We wish to thank K.Shima for encouragement, and our colleagues in
Tokyo University for their kind hospitality.

\end{document}